\documentclass[conference]{IEEEtran}
\usepackage{graphicx}
\ifCLASSINFOpdf
\else
\fi
\begin{document}
%
\title{ Community-based Immunization Strategies for Epidemic Control}



%
\author{\IEEEauthorblockN{Naveen Gupta\IEEEauthorrefmark{1},
Anurag Singh\IEEEauthorrefmark{1},
Hocine Cherifi\IEEEauthorrefmark{2}, 
\IEEEauthorblockA{\IEEEauthorrefmark{1}Computer Science \& Engineering Discipline,\\
PDPM Indian Institute of Information Technology, Design and Manufacturing Jabalpur, India\\
Email: naveen.gupta@iiitdmj.ac.in, anuragsg@iiitdmj.ac.in}
\IEEEauthorblockA{\IEEEauthorrefmark{2}University of Burgundy, LE2I UMR CNRS 6306, Dijon, France\\
Email: hocine.cherifi@u-bourgogne.fr}}}


\maketitle

\begin{abstract}
Understanding the epidemic dynamics, and finding out efficient techniques to control it, is a challenging issue. A lot of research has been done on targeted immunization strategies, exploiting various global network topological properties. However, in practice, information about the global structure of the contact network may not be available. Therefore, immunization strategies that can deal with a limited knowledge of the network structure are required. In this paper, we propose targeted immunization strategies that require information only at the community level. Results of our investigations on the SIR epidemiological model, using a realistic synthetic benchmark with controlled community structure, show that the community structure plays an important role in the epidemic dynamics. An extensive comparative evaluation demonstrates that the proposed strategies are as efficient as the most influential global centrality based immunization strategies, despite the fact that they use a limited amount of information. Furthermore, they outperform alternative local strategies, which are agnostic about the network structure, and make decisions based on random walks.
\end{abstract}



%
\IEEEpeerreviewmaketitle

\section{Introduction}

Outbreak of infectious diseases is a serious threat to the lives of people and it also brings serious economic loss for the victim countries. So, it is very important to discover the propagation rules in social groups in order to prevent the epidemics or at least to control the epidemic spreading. Vaccination allows to protect people and prevent them to transmit disease among their contacts. As mass vaccination is not always feasible, due to limited vaccination resources, targeted immunization strategies are of prime interest for public health. Impact of the contact network topology on disease transmission is a hot topic in the complex network litterature. A lot of work has been done towards this direction \cite{EpdInt,pastorepidemic,gong2013efficient,halloran2008modeling,barthelemyvelocity,singh2012rumour,Anuappb,anuncc}.

Top-ranked influential spreaders need to be identified for targeted immunization. In social networks, nodes with high centrality are considered to be the influential nodes. Since the most central nodes can diffuse their influence to the whole network faster than the rest of nodes, they are the ones to be targeted. Unfortunately, there is no general consensus on the definition of centrality and many measures have been proposed.  Degree centrality and betweenness centrality are the most popular ones \cite{degcent1,degcent2,Betcent1,Betcent2}. Targeting of high-centrality individuals in a network is a global strategy because it requires the knowledge of the whole network. The main drawback of these strategies is that very often there is no information available about the global structure of the real-world networks. Hence, efficient immunization strategies, based on locally available network information are of prime interest. Such strategies rely only on local information around selected nodes. The most basic strategy is random immunization where target nodes are picked at random regardless of the network topology. In acquaintance immunization, the selected node is chosen randomly among the neighbors of a randomly picked node. As randomly selected acquaintances of nodes are likely to have more connections than randomly selected nodes, this method targets highly connected nodes. It can be viewed as a local approximation of a global degree based strategy.  

It is well-recognized that numerous real-world networks ranging from biological to social systems share some topological properties. They are scale-free and exhibit the small-world property. Furthermore, they are characterized by a high clustering coefficient and a well-defined community structure. While a lot of research has been done to study and understand the effect of the degree distribution of the network on the dynamics of epidemic spreading, the impact of the community structure has not been paid a lot of attention.  More recently, this property of the networks is exploited to control the spreading of disease \cite{salathe, hebert}.

Unfortunately, these studies feature two significant shortcomings. Firstly, the community structure of the empirical data is questionable. Indeed in \cite{salathe} the ground truth community structure is  based on a functional definition that is not necessarily encoded in the network topology while in \cite{hebert} the authors rely on the results of a community detection method to uncover the community structure. As there is no consensual community detection method, results are very sensitive to this issue. Secondly, simulated data are based on simple models that do not reproduce properly the community structure observed in real-world networks. The main contributions of this paper are twofold. To overcome previous research drawbacks, we  analyze the impact of the community structure using a more realistic community-structured benchmark with controlled topological properties. Furthermore, we introduce new local influence measures based on the community structure. An extensive experimental evaluation  is conducted in order to investigate their efficiency as compared to global centrality measures and  the local Community Bridge Finder algorithm (CBF) introduced in \cite{salathe},  which has been designed for community structured networks.
 The remainder of this paper is organized as follows. In section 2, the benchmark model is introduced and its main properties are recalled. In the section 3, the immunization techniques are presented. Section 4 is devoted to the experimental results. We conclude with a discussion of our observations and findings in section 5.

\section{Synthetic Benchmark Data}
Generative models allow producing large collections of synthetic networks easily and quickly. These models also allow to control some topological properties of the generated networks, so as to get them close to the targeted system features. The only point of concern is how closely the  generated networks are able to represent the real-world networks, which is necessary for obtaining relevant test results. To date, the LFR (Lancichinetti, Fortunato and Radicchi)\cite{LFR} model is the most efficient solution in order to generate synthetic networks with community structure. Consequently, it is used to generate the networks with a non-overlapping community structure. It is based on the configuration model (CM) \cite{CM}, which generates networks with power law degree distributions. The generative process of the LFR algorithm performs in three steps. First, a network with power-law degree distribution with exponent $\gamma$, is generated using the configuration model. Second, virtual communities are defined so that their sizes follow a power-law distribution with exponent $\beta$. Each node is randomly assigned to a community, with the constraint that the community size must be greater than or equal to the node's internal degree. Third, an iterative process takes place in order to rewire certain links, so as to get the proportion of intra-community and inter-community links close to the mixing coefficient value $\mu$, while preserving the degree distribution. This model guarantees to obtain realistic features (power-law distributed degrees and community sizes) for the generated networks. It also includes a rich set of parameters that can be tuned to get the desired network topology. These parameters are, the mixing parameter $\mu$, the average degree $k$, the maximum degree $k_{max}$, the maximum community size $c_{max}$, and the minimum community size $c_{min}$. 
For small $\mu$ values, the communities are distinctly separated and thus easily identifiable because of lesser inter-community links. Whereas, when $\mu$ increases, the proportion of inter-community links becomes higher, making community distinction and identification a difficult task. The network has no community structure for a limit value of the mixing coefficient given by: 
\vspace{-0.9em}
\begin{equation}
\mu_{lim} > (n - n_c^{max})/n,
\end{equation}
where $n$ and $n_c^{max}$ are the number of nodes in the network and in the biggest community, respectively \cite{LF2009b}.

\section{Immunization Strategies}
Targeted immunization strategies can be divided in two categories based on their requirement about the knowledge of the network topology. Global strategies exploit the knowledge of the full network structure in order to find the influential nodes while local strategies are able to work with a limited amount of information.

\subsection{\textbf{Global strategies:}} These immunization strategies are based on an ordering of the nodes of the whole network according to an influence measure. Nodes are then targeted (removed) in the decreasing order of their rank. The influence of a node is computed according to some centrality measures. In this study, the most influential centrality measures (degree and betweenness) are considered in order to compare with the proposed strategies. We have not considered k-core centrality for evaluation as it is shown to be not very effective in finding the influential nodes for targeted immunization.\cite{hebert}

\textbf{1. \textit{global\_deg}:} Degree centrality denotes the number of immediate neighbors of a node, i.e. which are only one edge away from the node. It is simple but very coarse. It can be interpreted as the number of walk of length 1 starting at the considered node. It measures the local influence of a node. It has many ties and fails to take into account the influence weight of even the immediate neighbors. Even if it is a local measure, the immunization strategy is global because it needs to rank all the nodes of the network according to their degree. 

\textbf{2. \textit{global\_bet\_cent}:} Betweenness centrality defines the influence of a node based on the number of shortest paths between every pair of node, of which it is a part of. It basically tries to identify the influence of a node in terms of information flow through the network. In this strategy, the nodes are targeted based on their overall betweenness centrality. The computation of betweenness has high time complexity.

Note that  removing a node can modify the network topology and hence the centrality of the remaining nodes might not be the same. So, the centralities of the remaining nodes after removing the node with highest centrality need to be recalculated. In the global strategies, we have recalculated the centralities of the remaining nodes after removing nodes with highest respective centrality measures. 

\subsection{\textbf{Local strategies: }} Various methods based of the community characteristics are proposed to immunize or remove the nodes from the communities in order to control the epidemic spreading. These strategies are local because they only require information at the community level. In the case where the community structure is unknown, it can be uncovered with local community detection algorithms. Therefore, these strategies do not require any information about the global structure of the network. In a network with community structure, the degree of a node can be split into two contributions: the intra-community links connecting it to nodes in its own community and  the inter-community links connecting it to nodes outside its community. Strength of the community structure of a network depends upon inter and intra-community links. A network is said to have a strong and well-defined community structure if a small fraction of total links in the networks lies between the communities. On the contrary, if a large fraction of total links lies between the communities, then the network does not contain well-defined communities and it is said to have a weak community structure. 

The topology of a network can be fully specified by its adjacency matrix $A$. In the case of an undirected, unweighted network, $A(i, j)$ is equal to 1 if $i$ and $j$ are directly connected to each other otherwise it is equal to zero. Considering a community $C$ of a network, the total degree of a node $i$ can be split in to two parts: 
\begin{equation} 
k_i(C)  =  k_i^{in}(C) + k_i^{out}(C).
\end{equation}
The degree $k_i$ of a node $i$ is equal to the total number of its connections, $k_i = \sum_j  A(i ,j)$. The Indegree of a node is equal to the number of edges connecting it to other nodes of the same community and can be calculated as $k_i^{in}(C)  =  \sum_{j\in C } A(i, j)$. The Outdegree of a node $i$ is equal to the number of  connections to the nodes lying outside the community and can be calculated as $k_i^{out}(C)  =  \sum_{j \notin C}  A(i, j)$.

\textbf{1. \textit{inout\_diff\_nodes}:} In this strategy, the nodes targeted for immunization are ranked according to the difference of their indegree and outdegree. For a node $i$, this difference is given by:
\begin{equation}
k_i^{iod} = k_i^{in} -  k_i^{out}.
\end{equation}
A fraction of the nodes with highest $k_i^{iod}$ in each community are removed from the network. The reason for selecting these nodes is that they are more committed to their community and have very few connections outside their community. It is speculated that, removing these nodes will weaken the community structure of the network.

\textbf{2. \textit{outin\_diff\_nodes}:} In this case, nodes are targeted for immunization based on the difference of their outdegree and indegree. For a node $i$, it is given by: 
\begin{equation}
k_i^{oid} = k_i^{out} -  k_i^{in}.
\end{equation}
These nodes share more connections with nodes outside their community than inside. So, they act as bridge between the communities and are responsible for spreading the information outside of their community. It is speculated that removing such nodes will stop the information flow between different communities.

\textbf{3. \textit{indeg\_nodes}:} In this strategy, a given fraction of total nodes in each community with the highest indegree centrality are selected for immunization. These nodes can be considered as the core of their community which are responsible for maximum information flow inside the community. If these nodes are removed, it may weaken the community structure of the network and thus prevent the epidemic spreading inside the communities.

\textbf{4. \textit{outdeg\_nodes}:} In this strategy, nodes are targeted based on their outdegree centrality. A fraction of nodes with highest outdegree centrality, in each community are removed from the network with the intention of isolating the different communities. The idea is to break the bridges between the communities and thus to prevent the epidemic spreading across the communities.

\textbf{5. \textit{community bridge finder (CBF)}:} Along with degree and betweenness centrality based strategies, the proposed strategies are also compared with a stochastic strategy, \textit{CBF}, proposed by Salathe \textit{et al.} \cite{salathe}. The CBF algorithm is a random walk based algorithm, aimed at identifying nodes connected to multiple community. The algorithm begins with selecting a random node as the starting node. Then a random path is followed until a node is found that is not connected to more than one of the previously visited nodes on the random walk. It assumes that such node is more likely to belong to a different community. This strategy is completely agnostic about the network structure.

In the local strategies the number of nodes to be removed from a community is proportional to the community size. Hence, more nodes are removed from larger communities than from the smaller communities.

\vspace{0.9em}
\section{Experimental Results}
To investigate the efficiency of the proposed strategies, synthetic networks are generated using the LFR algorithm. Several studies of real-world networks are considered in order to select appropriate values for the network parameters. For the power-law exponents, we have used $\gamma = 3$ and $\beta = 2$, which seem to be close to most of the real-world networks. Concerning the number of nodes and links, no typical values emerge. Studies show that the size of real-world networks can differ very much, ranging from tens to millions of nodes. It is also difficult to characterize the average and maximal degrees of the network because of their variable nature. As a result, we selected some consensual values for these parameters, while also keeping in mind the computational aspect of the simulations \cite{Hocine}. The values of the parameters used for LFR networks generation in this study are given in the Table \ref{t1}.

\begin{table}[htb!]
\caption{Parameters for the LFR network generation} \label{t1}
\centering
\begin{tabular}{| c | c |} \hline
\hline
Number of nodes, n & 7500 \\ \hline
Average degree, $\langle k \rangle$ & 10 \\ \hline
Maximum degree, $k_{max}$ & 180 \\ \hline
Mixing parameter, $\mu$ & 0.3, 0.5, 0.7 \\ \hline
$\gamma$ & 3 \\ \hline
$\beta$ & 2 \\ \hline
Minimum community size, $C_{min}$ & 5 \\ \hline
Maximum community size, $C_{max}$ & 180 \\ \hline
\end{tabular}
\end{table}

In order to better understand the influence of the community structure, networks with various $\mu$ values (ranging from 0.3 to 0.7) are generated. For each $\mu$ value, 10 sample  networks are generated. To study the spread of an infectious disease in a contact network, the popular SIR model of epidemic spreading is used. Each node in the network is in one of the three possible states: (S)usceptible, (I)nfected, or (R)esistant/immune. Susceptible nodes represent the individuals which are not yet infected with the disease. Infected nodes are the ones which have been infected with the disease and can spread the disease to the susceptible nodes. Resistant nodes represent the individuals which have been infected and have been immunized or died. These nodes are not able to be infected again, neither can they transmit the infection to others. Initially, all the nodes are treated as susceptible. After this initial set-up, a fraction of  nodes are chosen at random to be infected and the remaining nodes are considered to be susceptible. The infection spreads through the contact network during a number of time steps, where $\lambda$ is the transmission rate of infection being spread from an infected node to a susceptible node, and $I$ is the number of infected neighboring nodes. Infected nodes recover with rate $\sigma$ at each time step. After the recovery of a node occurs, its state is toggled from infected to resistant. Simulations are halted if there are no infected nodes left in the network, and the total number of infected nodes is analyzed. Each simulation is run on these 10 networks and the mean values of the results together with their standard deviations are reported in the figures.

\begin{figure}[htb!]
\begin{center}
\includegraphics[width=0.9\linewidth, height=2.3in]{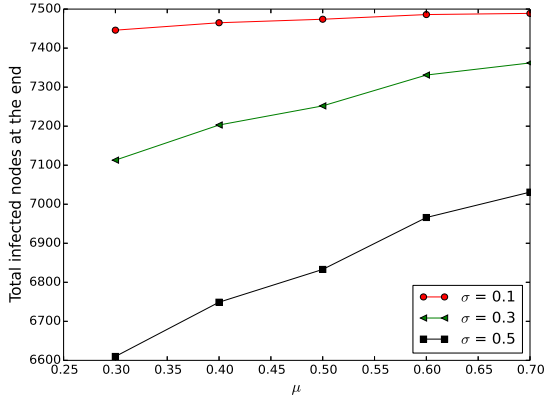}
\caption{Effect of $\mu$, mixing parameter of the LFR Network on the total infected nodes. Here, $\lambda$ = 1}
 \label{f1} 
 \vspace{-1.8em}
\end{center}
\end{figure}

Fig. \ref{f1} illustrates the impact of the mixing parameter, $\mu$ on the total infected nodes during the epidemic spreading. It can be observed that as $\mu$ increases, the total number of infected nodes during the epidemic spreading also increases. Indeed, the inter-community links i.e. the links between communities act as the carrier of the information across different communities, and consequently, they play a major role in the epidemic spreading. The infection starting inside a community spreads easily to other communities through these inter-community links. The total number of infected nodes increases when the community structure gets weaker. On the contrary, when the communities are very cohesive, there is few inter-community links and the epidemic is trapped in the community where it started.  The simulations are performed with different values of recovering rate, $\sigma$, as there might be different rates of recovery  in real world scenarios. It can be observed that for small values of $\sigma$, epidemic spreads over a larger population than in the case of large value of $\sigma$. Indeed, infected nodes recover at a very lower rate in the case of low $\sigma$, and hence keep on spreading the infection. It takes more time to get all the infected nodes in the network to the recovered state. Whereas, when the recovery rate is high, the infected nodes recover before the epidemic spreads to a larger population.

\begin{figure}[htb!]
\begin{center}
$\begin{array}{c}
\includegraphics[width=0.9\linewidth, height=2.4 in]{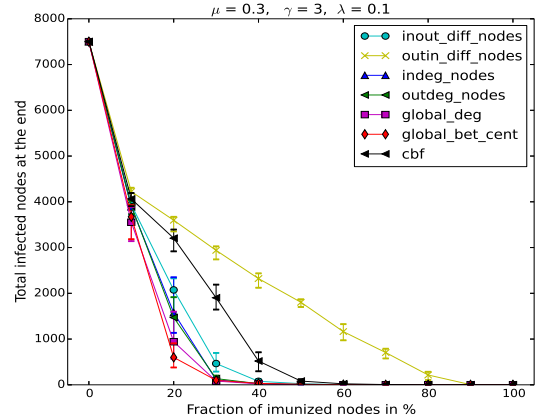}\\
\mbox{(a)}\\
\includegraphics[width=0.9\linewidth, height=2.4 in]{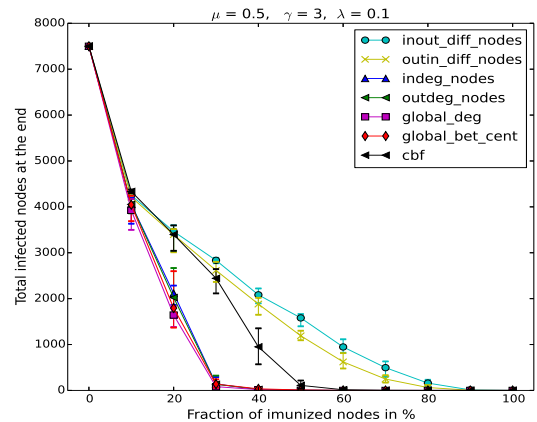}\\
\mbox{(b)}\\
\includegraphics[width=0.9\linewidth, height=2.4 in]{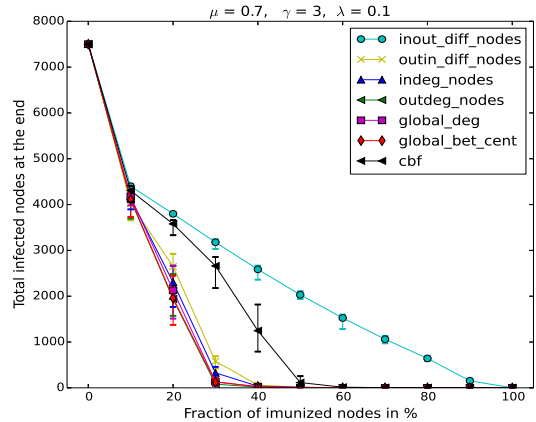} \\
\mbox{(c)} \\
\end{array}$
\end{center}
\caption{Effect of various immunization strategies on the total number of infected nodes during the SIR simulation on LFR network with $\sigma$ = 0.1, $\lambda$ = 0.1 and (a) $\mu$ = 0.3 (b)  $\mu$ = 0. 5 (c) $\mu$ = 0.7} \label{f2}
\vspace{-1.8em}
\end{figure}

\begin{figure}[t]
\begin{center}
$\begin{array}{c}
\includegraphics[width=0.9\linewidth, height=2.4 in]{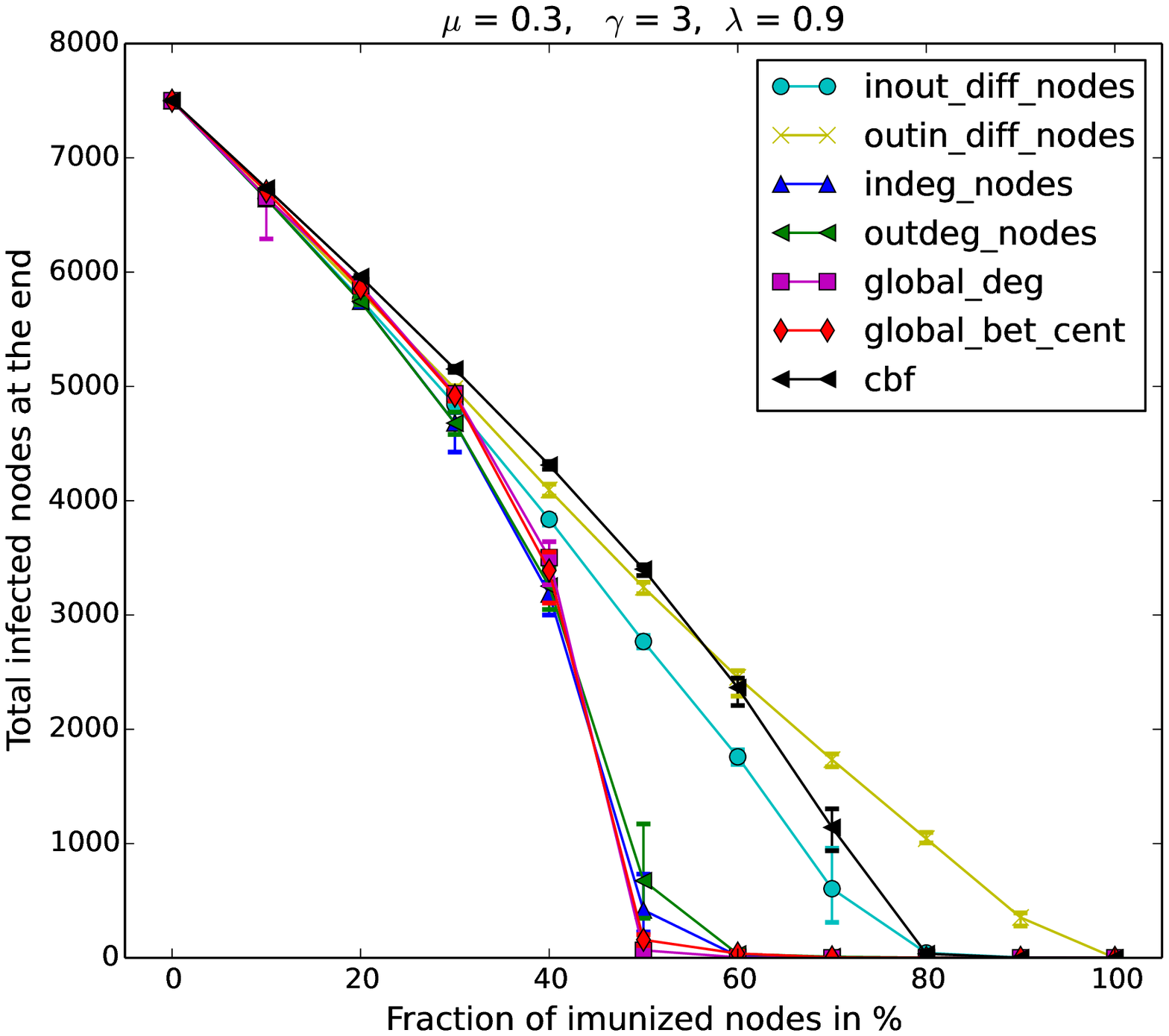}\\
\mbox{(a)}\\
\includegraphics[width=0.9\linewidth, height=2.4 in]{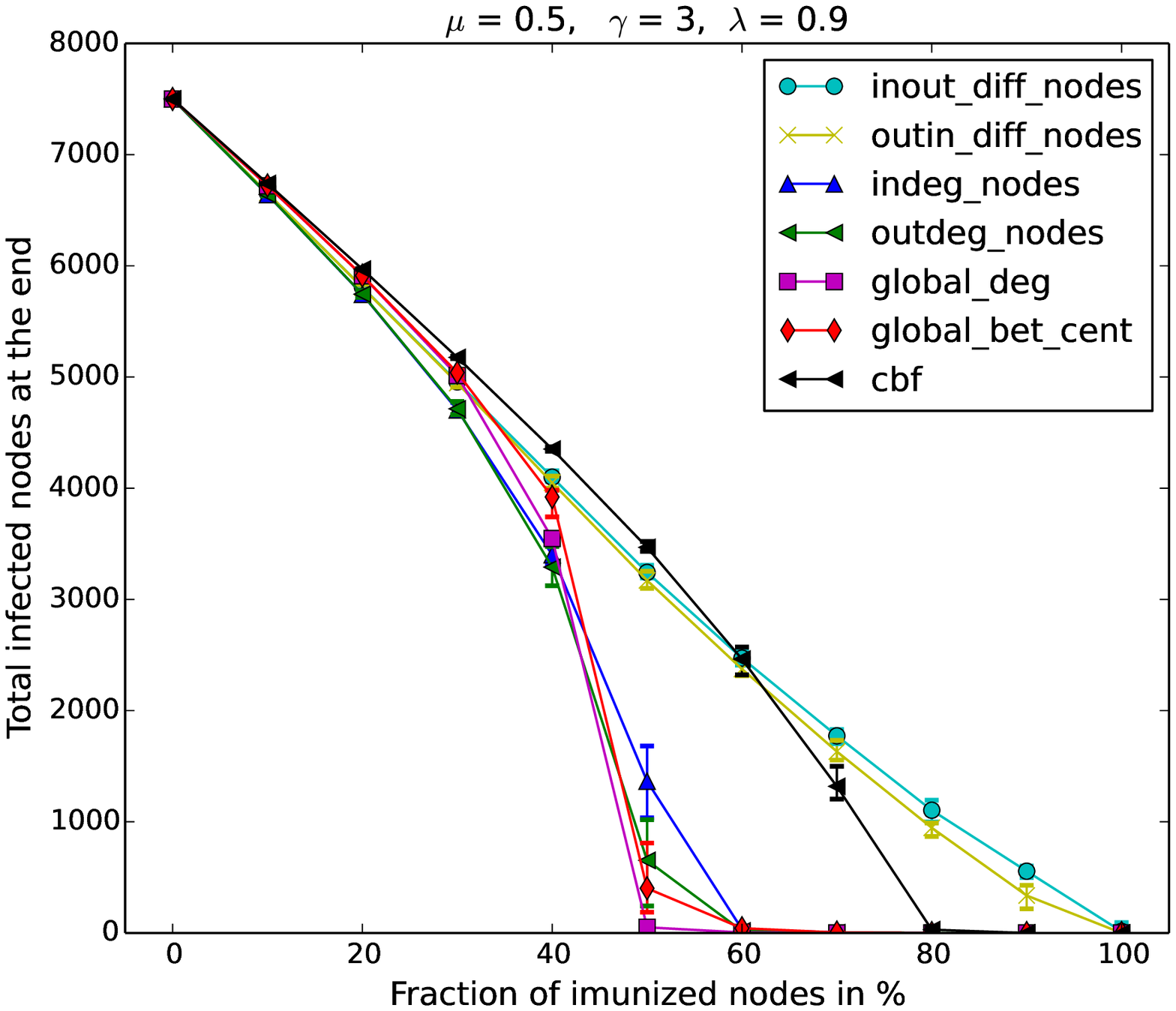}\\
\mbox{(b)}\\
\includegraphics[width=0.9\linewidth, height=2.4 in]{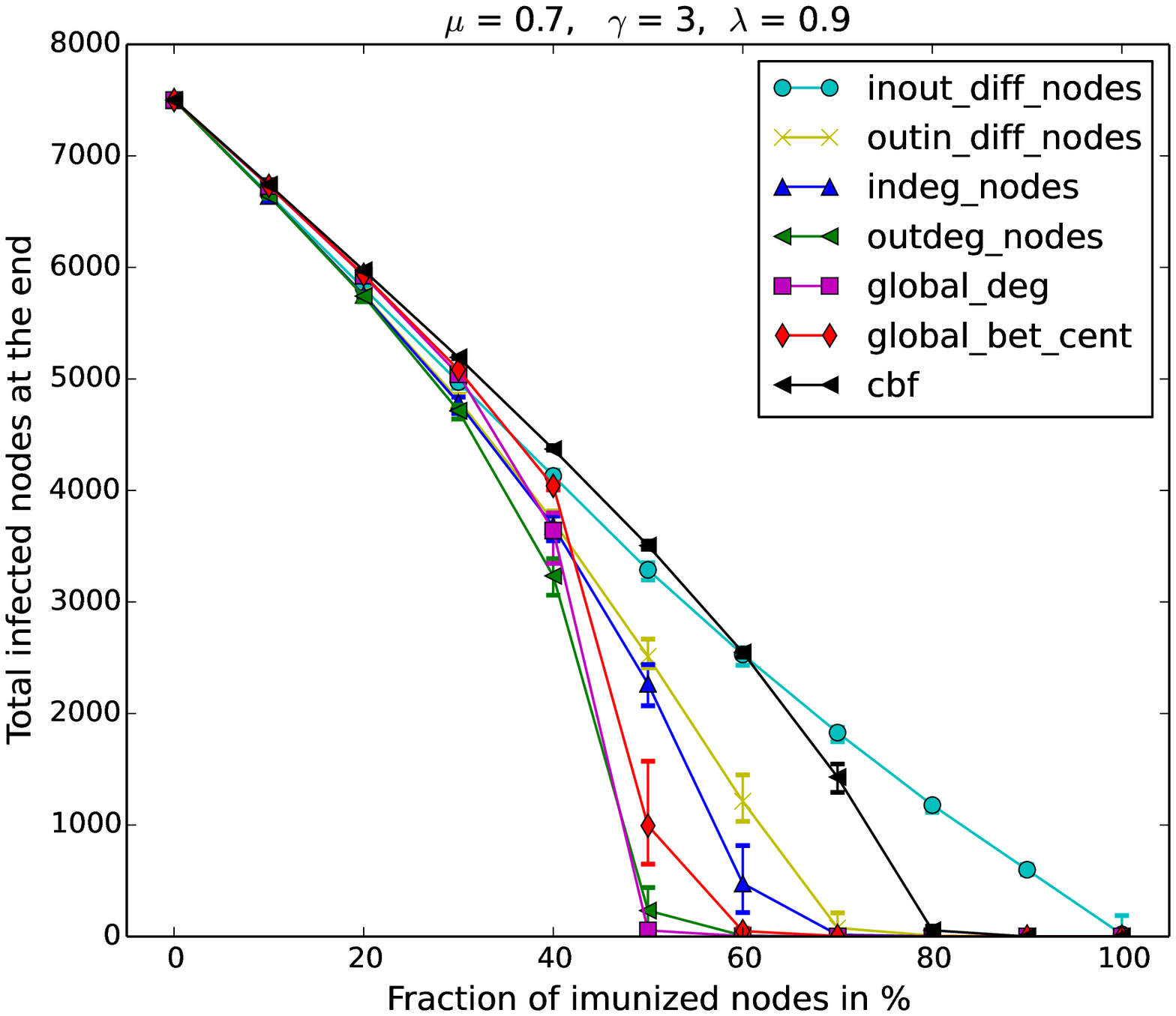}\\
\mbox{(c)}
\end{array}$
\end{center}
\caption{Effect of various immunization strategies on the total infected nodes during the SIR simulation on LFR network with $\sigma = 0.1$, $\lambda$ = 0.9 and (a) $\mu$ = 0.3 (b)  $\mu$ = 0. 5 (c) $\mu$ = 0.7} \label{f3} 
\vspace{-1.8em}
\end{figure}

The results of the simulations using various available and proposed immunization strategies are shown in Figs. \ref{f2} and \ref{f3}. Fig. \ref{f2} illustrates the results for  low values of transmission and recovering rate ($\sigma$ and $\lambda$ are both equal to 0.1). Note that the global centrality based methods (degree, betweenness) are very efficient. After removing or immunizing only 30\% of the nodes, epidemic spreading has died. This is expected as both the methods exploit the information about the overall network topology. When the \textit{CBF} strategy is used, 50\% nodes need to be removed in order to stop the epidemic spreading. This is the price to pay for not knowing the global structure of the network. More interesting, the proposed indegree and outdegree centrality based methods, \textit{indeg\_nodes} and \textit{outdeg\_nodes} respectively, are almost as effective as the global centrality methods despite the fact that they are agnostic about the full network structure. Furthermore, they perform a lot better than the \textit{CBF} strategy, with just the information about the community structure of the network. The reason why the proposed methods are as effective, can be explained by the specific position of the targeted nodes within the network. The nodes with the highest indegree can be considered as the core points of their community. In a network of people, these individuals are the leaders, representatives or agents of information flow in their communities. These high indegree nodes are connected to many other nodes in their community. Most of the regular nodes in a community are not directly connected to each other. They are connected with each other through the paths which would most likely contain these high indegree nodes. When these high indegree nodes are removed from the communities, the communities breaks  from inside. In other words, the paths connecting the regular nodes to each other are broken. The remaining nodes are not able to contact each other and thus the epidemic is not able to affect a significant part of community, and it dies soon. Whereas the nodes with the highest outdegree are the ones which have lot of connections to other communities. In a group of people, these are the individuals which pass all the information contained in their group to other groups. These nodes can be considered to be the bridges between the communities. When these nodes are removed, most of the paths or bridges between different communities are lost. Communities are isolated and thus the epidemic is not able to spread across the communities. The underlying idea in selecting these centrality measures is that they intuitively represent the global degree and betweenness centralities at the community level. A node with a high indegree or outdegree has generally a high overall degree. Indeed, the total degree of a node is the addition of indegree and outdegree. A node with high indegree is probably a node with a high betweenness measure in its community, as it will be contained in most of the paths connecting the regular nodes to each other in the community. Similarly, a node with high outdegree is probably a node with high betweenness measure in the overall network. Indeed, these high outdegree nodes are part of most of the paths connecting the nodes falling in different communities.

Fig. \ref{f2}(a) shows the results of the immunization strategies for a strong community structure with $\mu$ = 0.3. In other words, the communities are densely connected and there is a low proportion of inter-community links. Here, we observe that even the proposed \textit{inout\_diff\_nodes} method gives better results than the \textit{CBF} strategy. In the case of strong community structure, most of the nodes have all the connections inside their community. The \textit{inout\_diff\_nodes} method selects the nodes to be immunized on the basis of the highest difference of indegree and outdegree. It targets the nodes which have more connections inside their community than outside. These nodes are more committed to their community, and have very few connections outside of their community. These nodes are the ones which once infected from outside, have a greater chance of infecting the whole community. So, targeting these nodes helps to prevent the epidemic spreading in a network with a strong community structure. In Fig. \ref{f2}(c), $\mu$ = 0.7, the network has a weak community structure, i.e. around 70\% of the links lie between the communities. In this case, \textit{inout\_diff\_nodes} method is not that effective, but \textit{outin\_diff\_nodes} method works better than both \textit{CBF} and \textit{inout\_diff\_nodes}. \textit{outin\_diff\_nodes} method selects the nodes on the basis of highest difference of their outdegree and indegree. These are the nodes which have more connections to the other communities than to their own community. After catching the infection from inside their community, these nodes quickly spread the infection to other communities because of their numerous outside connections. In the case of weak community structure, the \textit{outin\_diff\_nodes} method works better as it targets the nodes which work as bridges between the communities and are responsible for spreading the epidemic across different communities. In the case of $\mu$ = 0.5, (See Fig. \ref{f2}(b)) when inter and intra-community links are almost as numerous, the \textit{inout\_diff\_nodes} and \textit{outin\_diff\_nodes} are not very effective. Indeed, the difference of indegrees and outdegrees of the nodes is around zero in this case, and these methods are not be able to decide which are the influential nodes.

Fig. \ref{f3} reports the SIR simulation results for $\lambda$ = 0.9 and $\sigma$ = 0.1. The results are very similar except the fact that in this case, a greater number of nodes need to be immunized in order to stop the epidemic spreading. This is true for all the strategies. For example, when $\lambda$ is equal to 0.1, degree and betweenness centrality based methods required only 30\% of the nodes to be removed to mitigate the epidemic spreading, whereas now they require 50\% of the nodes to be immunized or removed. Indeed, when $\lambda$ increases, the probability of an infected node to infect its neighbors gets higher. So, the epidemic spread at a higher rate, and thus more nodes are needed to be immunized to prevent the epidemic spreading.

To summarize, according to the experimental results, the proposed indegree and outdegree centrality based strategies- \textit{indeg\_nodes} and \textit{outdeg\_nodes} respectively, are effective in identifying the influential nodes to be selected for immunization to prevent or mitigate the epidemic spreading. These methods are as effective as the global degree and centrality based methods, but they do not require any information about the global structure of the network. The proposed strategies perform better than the \textit{CBF} algorithm, with only the information about the local (community-wise) structure of the network. This suggests that the local information is sufficient in order to design an immunization strategy. 

\section{Conclusions}
There might not be enough information available about the global structure of the underlying relevant contact network in order to control the epidemic spreading. Therefore, efficient immunization strategies are required that can work with the information available at the community level. Results of our investigation, on a realistic synthetic benchmark, show that the community structure plays a major role in the epidemic dynamics. It is observed that the proposed indegree and outdegree centrality based immunization strategies are efficient methods to control the epidemic spreading. These strategies work as well as the global centrality measures (degree and betweenness), without any knowledge of the global network structure.  Therefore, these centrality measures defined at the community level are good approximations of global centrality measures. The indegree  of a node represents its degree and betweenness centralities, relative to its own community, whereas, the outdegree centrality of a node represents these global centralities relative to the other communities of the network. However, unlike global centrality measures, the proposed measures  do not need to have any knowledge of the global network topology and thus can be easily and quickly computed. Furthermore,they perform better than \textit{CBF}. Performances of the two other proposed local strategies (\textit{inout\_diff\_nodes} and \textit{outin\_diff\_nodes}) depends on the strength of the community structure. In a network with strong community structure, \textit{inout\_diff\_nodes} method performs better, whereas \textit{outin\_diff\_nodes} method is more efficient in the case of weak community structure. Finally, the main lesson of this work is that exploiting the local information on the network topology can be very effective in order to design  efficient immunization strategies that can be used in large scale networks. These preliminary results pave the way for more investigations on  alternative community topological measures.

\section*{Acknowledgement}
The authors thank Mr. Upendra Singh (B.Tech., MANIT Bhopal, India) for implementing the SIR model of epidemics, used in this paper.

\bibliographystyle{IEEEtran}
\bibliography{Ref1}

\begin{thebibliography}{10}
\providecommand{\url}[1]{#1}
\csname url@samestyle\endcsname
\providecommand{\newblock}{\relax}
\providecommand{\bibinfo}[2]{#2}
\providecommand{\BIBentrySTDinterwordspacing}{\spaceskip=0pt\relax}
\providecommand{\BIBentryALTinterwordstretchfactor}{4}
\providecommand{\BIBentryALTinterwordspacing}{\spaceskip=\fontdimen2\font plus
\BIBentryALTinterwordstretchfactor\fontdimen3\font minus
  \fontdimen4\font\relax}
\providecommand{\BIBforeignlanguage}[2]{{%
\expandafter\ifx\csname l@#1\endcsname\relax
\typeout{** WARNING: IEEEtran.bst: No hyphenation pattern has been}%
\typeout{** loaded for the language `#1'. Using the pattern for}%
\typeout{** the default language instead.}%
\else
\language=\csname l@#1\endcsname
\fi
#2}}
\providecommand{\BIBdecl}{\relax}
\BIBdecl

\bibitem{EpdInt}
S.~Boccaletti, V.~Latora, Y.~Moreno, M.~Chavez, and D.-U. Hwang, ``Complex
  networks: Structure and dynamics,'' \emph{Physics reports}, vol. 424, no.~4,
  pp. 175--308, 2006.

\bibitem{pastorepidemic}
R.~Pastor-Satorras and A.~Vespignani, ``Epidemic spreading in scale-free
  networks,'' \emph{Physical review letters}, vol.~86, no.~14, p. 3200, 2001.

\bibitem{gong2013efficient}
K.~Gong, M.~Tang, P.~M. Hui, H.~F. Zhang, D.~Younghae, and Y.-C. Lai, ``An
  efficient immunization strategy for community networks,'' \emph{PloS one},
  vol.~8, no.~12, p. e83489, 2013.

\bibitem{halloran2008modeling}
M.~E. Halloran, N.~M. Ferguson, S.~Eubank, I.~M. Longini, D.~A. Cummings,
  B.~Lewis, S.~Xu, C.~Fraser, A.~Vullikanti, T.~C. Germann \emph{et~al.},
  ``Modeling targeted layered containment of an influenza pandemic in the
  united states,'' \emph{Proceedings of the National Academy of Sciences}, vol.
  105, no.~12, pp. 4639--4644, 2008.

\bibitem{barthelemyvelocity}
M.~Barth{\'e}lemy, A.~Barrat, R.~Pastor-Satorras, and A.~Vespignani, ``Velocity
  and hierarchical spread of epidemic outbreaks in scale-free networks,''
  \emph{Physical Review Letters}, vol.~92, no.~17, p. 178701, 2004.

\bibitem{singh2012rumour}
A.~Singh and Y.~N. Singh, ``Rumor spreading and inoculation of nodes in complex
  networks,'' in \emph{Proceedings of the 21st international conference
  companion on World Wide Web}.\hskip 1em plus 0.5em minus 0.4em\relax ACM,
  2012, pp. 675--678.

\bibitem{Anuappb}
------, ``Nonlinear spread of rumor and inoculation strategies in the nodes
  with degree dependent tie stregth in complex networks,'' \emph{Acta Physica
  Polonica B}, vol.~44, no.~1, pp. 5--28, Jan 2013.

\bibitem{anuncc}
------, ``Rumor dynamics with inoculations for correlated scale free
  networks,'' in \emph{Communications (NCC), 2013 National Conference on},
  2013, pp. 1--5.

\bibitem{degcent1}
R.~Pastor-Satorras and A.~Vespignani, ``Immunization of complex networks,''
  \emph{Phys. Rev. E}, vol.~65, p. 036104, Feb 2002.

\bibitem{degcent2}
R.~M. Christley, G.~L. Pinchbeck, R.~G. Bowers, D.~Clancy, N.~P. French,
  R.~Bennett, and J.~Turner, ``Infection in social networks: Using network
  analysis to identify high-risk individuals,'' vol. 162, no.~10, pp.
  1024--1031, 2005.

\bibitem{Betcent1}
M.~J. Newman, ``A measure of betweenness centrality based on random walks,''
  \emph{Social Networks}, vol.~27, no.~1, pp. 39 -- 54, 2005.

\bibitem{Betcent2}
Z.~Hai-Feng, L.~Ke-Zan, F.~Xin-Chu, and W.~Bing-Hong, ``An efficient control
  strategy of epidemic spreading on scale-free networks,'' \emph{Chinese
  Physics Letters}, vol.~26, no.~6, p. 068901, 2009.

\bibitem{salathe}
M.~Salath{\'e} and J.~H. Jones, ``Dynamics and control of diseases in networks
  with community structure,'' \emph{PLoS computational biology}, vol.~6, no.~4,
  p. e1000736, 2010.

\bibitem{hebert}
L.~H{\'e}bert-Dufresne, A.~Allard, J.-G. Young, and L.~J. Dub{\'e}, ``Global
  efficiency of local immunization on complex networks,'' \emph{Scientific
  reports}, vol.~3, 2013.

\bibitem{LFR}
A.~Lancichinetti, S.~Fortunato, and F.~Radicchi, ``Benchmark graphs for testing
  community detection algorithms,'' \emph{Phys. Rev. E}, vol.~78, p. 046110,
  Oct 2008.

\bibitem{CM}
M.~Molloy and B.~Reed, ``A critical point for random graphs with a given degree
  sequence,'' \emph{Random structures \& algorithms}, vol.~6, no. 2-3, pp.
  161--180, 1995.

\bibitem{LF2009b}
A.~Lancichinetti and S.~Fortunato, ``Community detection algorithms: a
  comparative analysis,'' \emph{Physical review E}, vol.~80, no.~5, p. 056117,
  2009.

\bibitem{Hocine}
G.~K. Orman, V.~Labatut, and H.~Cherifi, ``Comparative evaluation of community
  detection algorithms: a topological approach,'' \emph{Journal of Statistical
  Mechanics: Theory and Experiment}, vol. 2012, no.~08, p. P08001, 2012.

\end{thebibliography}

\end{document}